\pdfoutput=1
\documentclass[%
reprint,
superscriptaddress,
twocolumn,
amsmath,amssymb,
aps,
pra,
]{revtex4-2}
\usepackage{footmisc}
\usepackage{xcolor}
\usepackage{graphicx}
\usepackage{dcolumn}
\usepackage{bm}
\usepackage{romannum}
\usepackage{siunitx}

\usepackage{titlesec}
\titleformat{\section}{\raggedright\fontsize{12.5}{25}\bfseries}{\arabic{section}.}{1em}{}
\pdfoutput=1

\DeclareUnicodeCharacter{0301}{\'a}
\DeclareUnicodeCharacter{00D6}{\"o}
\DeclareUnicodeCharacter{00C4}{\"a}
\DeclareUnicodeCharacter{00FC}{\"u}
\DeclareUnicodeCharacter{00DF}{\ss}

\usepackage{hyperref}

\usepackage{ulem}

\begin{document}
\pagenumbering{arabic}

\title{\fontsize{15}{19}\selectfont Resonance-Induced Anomalies in Temperature-Dependent Raman Scattering of PdSe$_{2}$}

\author{Omar Abdul-Aziz}
\affiliation{Universität zu Köln, II. Physikalisches Institut, Z\"ulpicher Straße 77, Köln D-50937, Germany}

\author{Daniel Wolverson}
\affiliation{Department of Physics and Centre for Photonics and Photonic Materials, University of Bath, BA2 7AY Bath, UK}

\author{Charles Sayers}
\affiliation{Dipartimento di Fisica, Politecnico di Milano, 20133 Milan, Italy}

\author{Ettore Carpene}
\affiliation{CNR-IFN, Dipartimento di Fisica, Politecnico di Milano, 20133 Milan, Italy}

\author{Fulvio Parmigiani}
\affiliation{Dipartimento di Fisica, Università di Trieste, via A. Valerio 2, 34127, Trieste, Italy}

\author{Hamoon Hedayat}
\email{hedayat@ph2.uni-koeln.de}
\affiliation{Universität zu Köln, II. Physikalisches Institut, Z\"ulpicher Straße 77, Köln D-50937, Germany}

\author{Paul H. M. van Loosdrecht}
\email{pvl@ph2.uni-koeln.de}
\affiliation{Universität zu Köln, II. Physikalisches Institut, Z\"ulpicher Straße 77, Köln D-50937, Germany}

\begin{abstract}
We report a comprehensive Raman study of the phonon behaviour in PdSe$_2$ in the temperature range of 5 K to 300 K. A remarkable change in the Raman spectrum is observed at 120 K: a significant enhancement of the out-of-plane phonon A$^{1}_{g}$ mode, accompanied by a suppression of the in-plane A$^{2}_{g}$ and B$^{2}_{1g}$ modes. This intriguing behavior is attributed to a temperature-dependent resonant excitation effect. The results are supported by density functional theory (DFT) calculations, which demonstrate that the electron-phonon coupling for the phonon modes varies and is strongly associated with the relevant electronic states. Furthermore, nonlinear frequency shifts are identified in all modes, indicating the decay of an optical phonon into multiple acoustic phonons. The study of Raman emission reported here, complemented by linear optical spectroscopy, reveals an unexpected scenario for the vibrational properties of PdSe$_2$ that holds substantial promise for future applications in PdSe$_2$-based optoelectronics.
\end{abstract}

\maketitle

\section{INTRODUCTION\label{Introduction}}
Transition metal dichalcogenides (TMDs) are a prominent category of van der Waals layered materials with outstanding optical and electronic properties and diverse correlated phases \cite{PhysRevB.26.6554, wang2012electronics, lopez2013ultrasensitive, wilson1969transition}. Among them, $\si{PdSe_2}$ has emerged as a material of particular interest due to its unique low-symmetry puckered atomic structure which in the literature is sometimes referred to as pentagonal \cite{zhang2015penta} and is characterized by an orthorhombic lattice of $\si{Pbca}$ space-group symmetry and $\si{D_{2h}}$ point-group symmetry \cite{gronvold1957crystal}. The distinctive bond arrangement within $\si{PdSe_{2}}$ leads to substantial interlayer interactions, resulting in significant layer-dependent electronic properties and anisotropic features \cite{deng2018strain, zhang2015penta}. Compared to conventional TMDs, $\si{PdSe_{2}}$ exhibits excellent stability in air, a unique linear dichroism conversion phenomenon, a high carrier mobility of $\sim$ 150 $\si{cm^2}\cdot\si{V^{-1}}\cdot\si{s^{-1}}$, and an exceptional long-wavelength infrared photoresponsivity \cite{oyedele2017pdse2, long2019palladium, yu2020direct, gu2020two}. $\si{PdSe_{2}}$ is a semiconductor with a bandgap in the infrared (IR) spectral region that is tunable from 0 to 1.3 eV when transitioning from its bulk to monolayer form\cite{oyedele2017pdse2, wei2022layer}.  Additionally, a sign change occurs in magnetoresistance at approximately 100~K, attributed to electronic anisotropy, despite $\si{PdSe_2}$ being a non-magnetic material \cite{zhu2021observation}. These unique features make $\si{PdSe_2}$ a promising candidate for advanced polarization-sensitive photodetectors and other optoelectronic applications \cite{wang2021applications,oyedele2017pdse2, zeng2019multilayered, wang2020noble, li2021phonon}.\\
\begin{figure*} 
\includegraphics[width=0.8\textwidth]{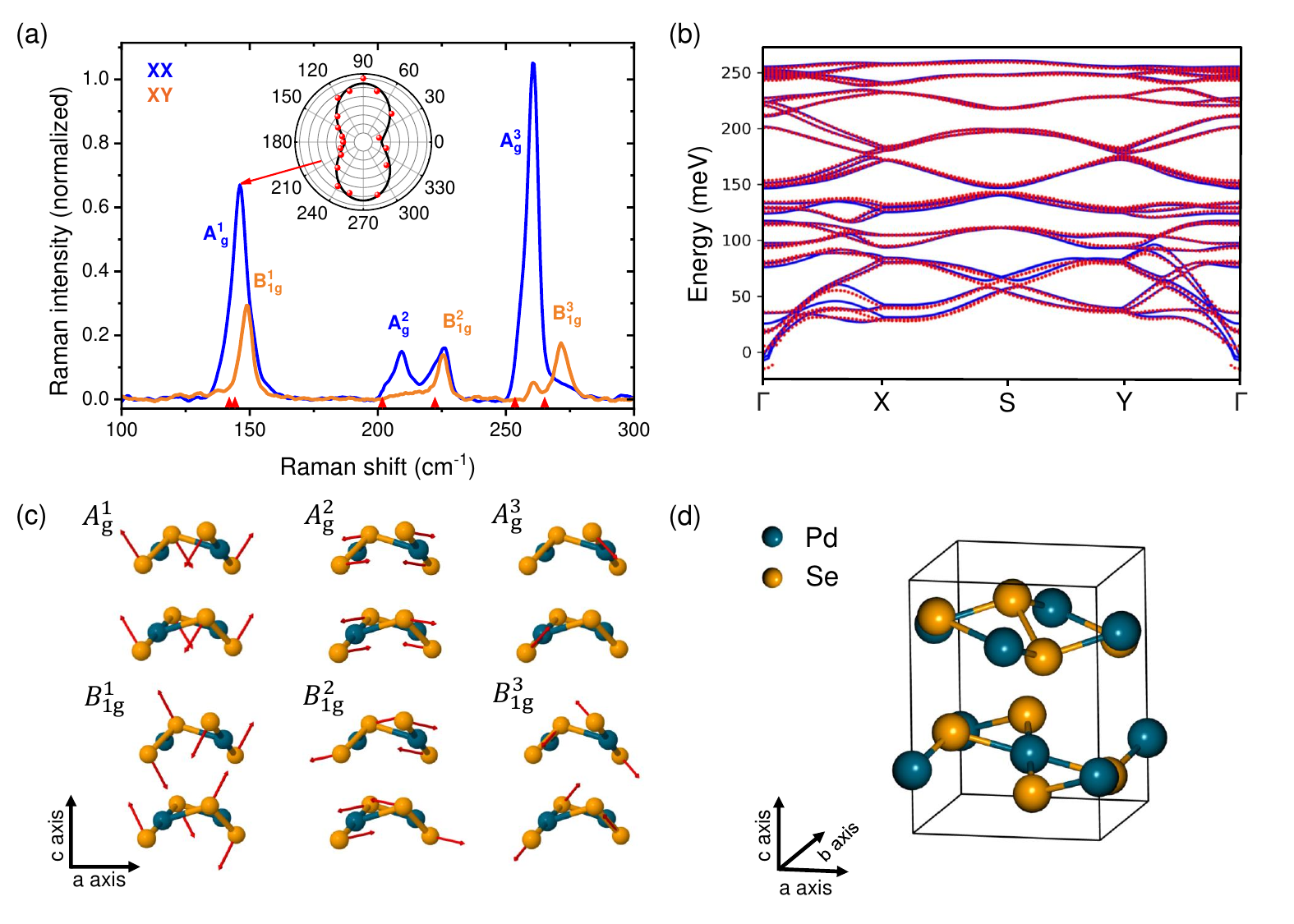}
\caption{(a) Raman spectra of bulk PdSe$_2$ in parallel (blue line) and cross configurations (yellow line) at room temperature. Phonon modes are labeled with an irreducible representation. The inset shows the Polarization-dependence of the Raman intensity of $\si{A_{g}^1}$ under parallel configurations (scattered dots). The fit of the Raman intensities versus polarization angles (black line). The red arrows indicate the phonon frequencies calculated by DFT. (b) DFPT calculation of the phonon dispersion of bulk PdSe$_2$ (blue) compared to the dispersion calculated via the EPW code used to obtain the electron-phonon coupling (red). (c) Atomic displacements corresponding to the phonon modes of PdSe$_2$, as depicted, projected along the b-axis and showing two layers, to indicate the phase relationship between layers. (d) The crystal structure of PdSe$_2$ unit cell. }
\label{Fig1}
\end{figure*}
Raman spectroscopy has been recently used to characterize the basic vibrational properties of $\si{PdSe_2}$ \cite{oyedele2017pdse2,yu2020direct}, however, a deeper understanding of the intricate electron-phonon interactions remains an area that deserves further exploration. For example, limited attention has been paid to the temperature-dependent behavior of Raman spectra in $\si{PdSe_2}$, with previous studies focusing primarily on temperatures above 300 K \cite{pi2021highly, li2021fast}. Despite demonstrating the anisotropy of lattice vibrations in $\si{PdSe_2}$ \cite{puretzky2018anomalous,luo2020anisotropic}, its connection to electronic states requires further clarification. Electron-phonon interactions within TMDs similar to other anisotropic-layered materials, such as black phosphorus \cite{zhong2022intrinsic, wu2015identifying, ribeiro2018raman}, give a strong Raman response. Therefore, Raman scattering is used as a powerful probe to gain insight into phonons and the way they interact with interband electronic transitions, uncovering the symmetry-dependent electron-phonon coupling of materials with anisotropic properties \cite{zhao2021computational,del2016atypical, tan2021breakdown, lai2021detection, del2014excited}. In these cases, the intensities of the phonon modes exhibit energy-dependent variations. A prototypical example is MoS$_2$, where the A$_{g}$ mode shows an enhancement when the excitation energy coincides with the exciton states A and B \cite{carvalho2015symmetry, nam2015excitation}. This enhancement has been attributed to resonance-induced symmetry breaking, as strongly localized exciton wave functions effectively break the symmetry of the system, thereby activating normally Raman-inactive modes. This selective increase in the intensity of phonon modes has been linked to symmetry-dependent electron-phonon coupling, as demonstrated in resonance Raman studies of MoS$_2$ using different excitation energies \cite{carvalho2015symmetry}. Similar to other TMDs, $\si{PdSe_{2}}$ also exhibits a selective response to different excitation energies as recently shown by Luo et al. \cite{luo2022excitation}.\\  
In this study, we present the Raman spectra of  $\si{PdSe_{2}}$  using different polarization configurations. Furthermore, we offer a comprehensive analysis of the evolution of the Raman spectrum in the temperature range of 5 K to 300 K. We find that the Raman intensity of the $\si{A^1_{g}}$ out-of-plane phonon modes experiences a significant enhancement at 120 K when excited with 2.33 eV, while the in-plane $\si{A^2_{g}}$ and $\si{B^2_{1g}}$ phonon modes are suppressed under the same conditions. This intriguing phenomenon can be understood as a result of resonance effects combined with diverse electron-phonon coupling of different modes, which were further confirmed by optical spectroscopy experiments. To gain a deeper understanding of the electron-phonon interactions responsible for these observations, we employed density functional theory (DFT) calculations. The calculations reveal details of how each phonon mode is coupled to the specific electronic orbitals. The results elucidate the intricate interplay between electronic states and phonons in $\si{PdSe_{2}}$, unveiling distinct electron-photon interactions associated with specific electronic states and phonon modes. Additionally, the study highlights phonon-phonon interactions spanning a wide temperature range that can be explained by a physical model that includes anharmonic contributions and offers an in-depth analysis of the nonlinear temperature-dependent Raman shifts of prominent optical phonon modes in $\si{PdSe_{2}}$. Such insights lay the groundwork for future developments in $\si{PdSe_{2}}$-based thermo-optoelectronic applications.

\begin{figure*}
\includegraphics[width=0.8\textwidth]{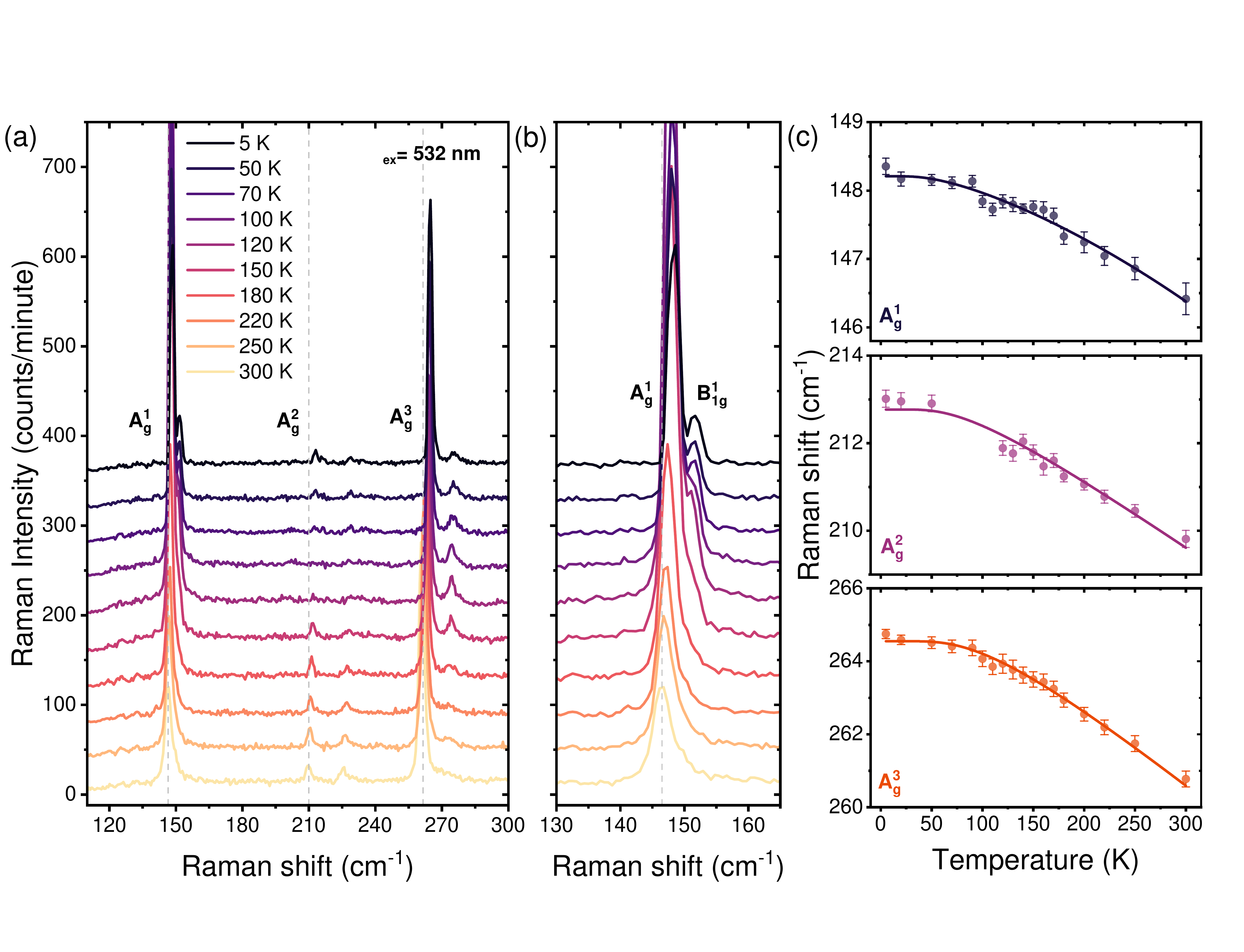}
\caption{(a, b) Temperature-dependent Raman spectra of bulk $\si{PdSe_{2}}$ under parallel configuration (xx), with an excitation energy of 2.33 eV (532 nm). (c) Raman frequencies as a function of temperature for the $\mathrm{A^1_{g}}$, $\mathrm{A^2_{g}}$, and $\mathrm{A^3_{g}}$ phonon modes. In each panel, we fit the data to the anharmonic model characterized by the processes of optical phonon decay (solid lines) using Eq.(1) over the entire temperature range 5 to 300 K.}
\label{Fig2}
\end{figure*}

\begin{figure}
\includegraphics[width=\columnwidth]{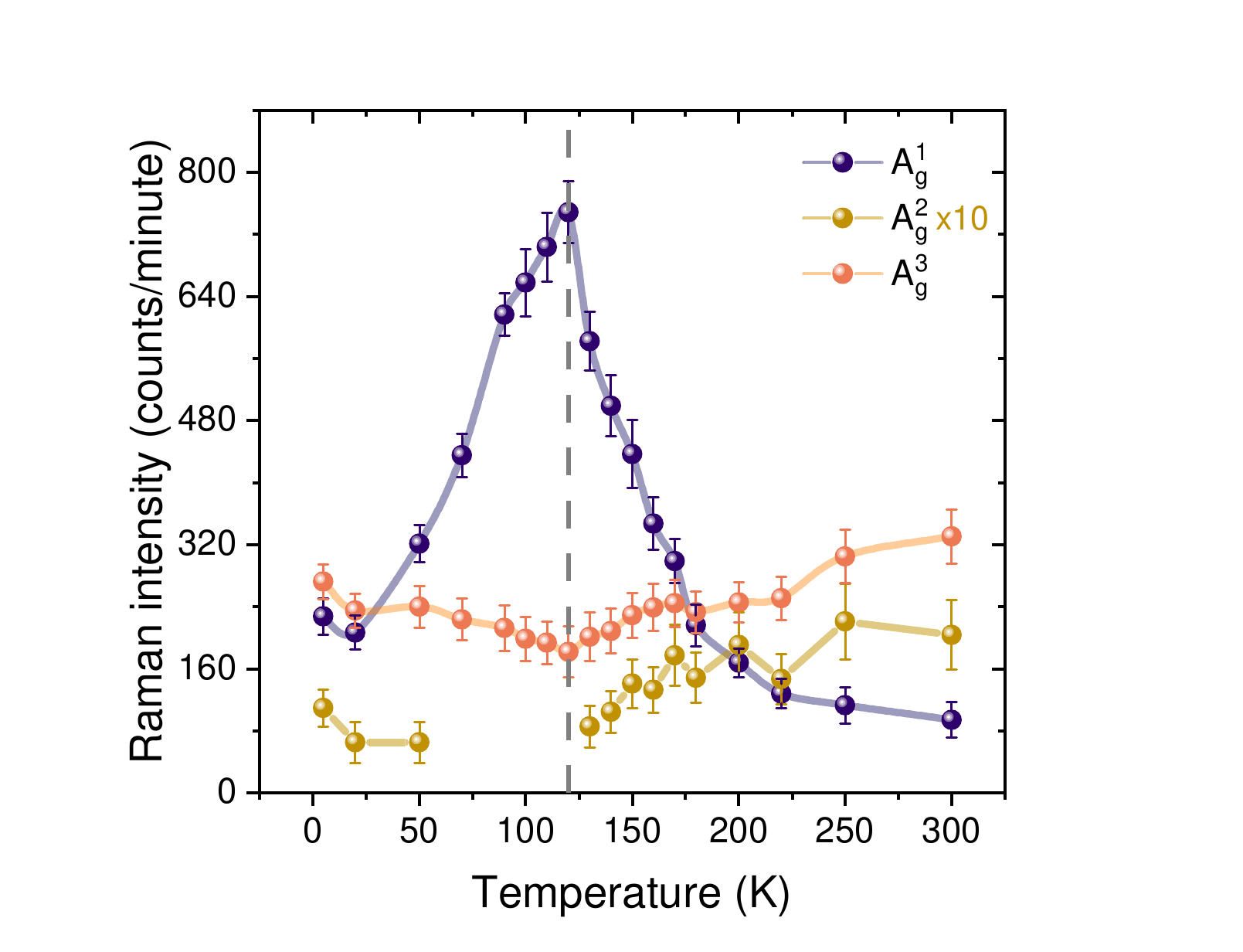}
\caption{The Stokes Raman spectral intensities, as a function of temperature, of the $\si{A^1_{g}}$, $\si{A^2_{g}}$, and $\si{A^3_{g}}$ modes. From 125 K down to 70 K, the $\si{A^2_{g}}$ mode exhibits zero intensity due to complete suppression. The vertical dashed line indicates 120 K. }
\label{Fig3A}
\end{figure}
\begin{figure*} 
\includegraphics[width=0.8\textwidth]{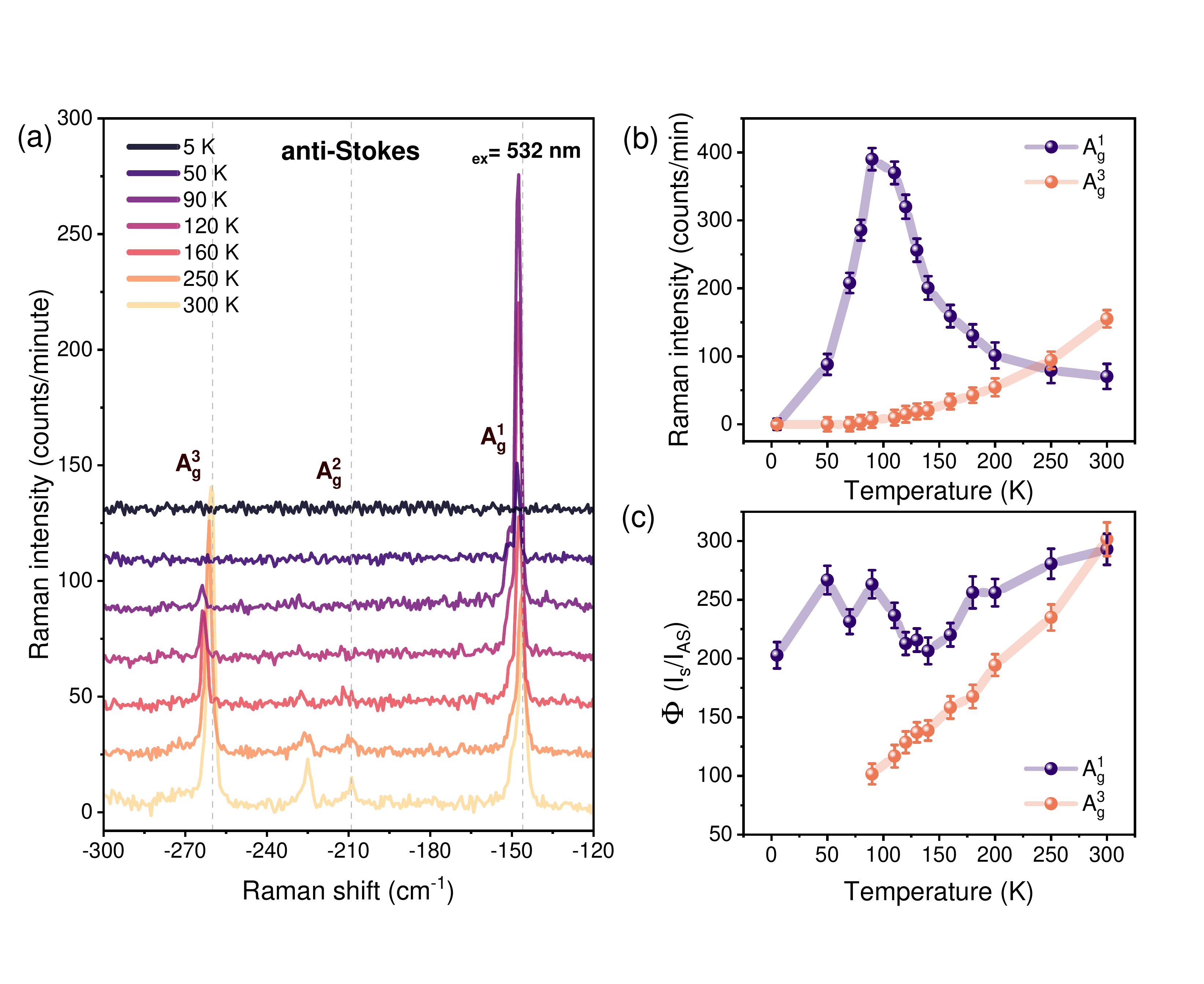}
\caption{(a) Temperature-dependent Raman anti-Stokes spectra of bulk $\si{PdSe_{2}}$ under parallel configuration (xx) with an excitation energy of 2.33 eV (532 nm). (b) Anti-Stokes Raman spectral weight intensities, as a function of temperature, of the $\si{A^1_{g}}$, and $\si{A^3_{g}}$ modes. (c) The calculated phonon temperature $\phi\left(I_{S} / I_{AS}\right)$ under non-resonance conditions for $\si{A^1_{g}}$ and $\si{A^3_{g}}$ phonon modes plotted as a function of temperature.}
\label{Fig3}
\end{figure*}
\section{RESULTS and DISCUSSIONS }\label{results}
Figure \ref{Fig1}(a), shows the comparative room temperature Raman spectra of bulk PdSe$_2$ using 532 nm (2.33 eV) laser excitation. The spectra were acquired in backscattering geometry with parallel (xx) and cross (xy) polarization configurations. We observe six characteristic Raman peaks corresponding to $\mathrm{A_g}$-symmetry modes at 146.7, 209.4, and 260.6 cm$^{-1}$, and $\mathrm{B_{1g}}$-symmetry modes found at at 148.7, 225.9, and 271.5 cm$^{-1}$. To validate the understanding of phonons in bulk PdSe$_2$, we performed calculations using density functional perturbation theory (DFPT) \cite{RevModPhys.73.515}. The phonon dispersion curves of bulk PdSe$_2$ along the high-symmetry crystallographic axes are presented in Figure \ref{Fig1}(b), while Figure \ref{Fig1}(c)  presents the calculated atomic displacements corresponding to the observed Raman modes. Figure \ref{Fig1}(b) also demonstrates that the phonon dispersion used in calculating the electron-phonon coupling via EPW \cite{noffsinger2010epw} (discussed in the Supplementary
Information) is consistent with the DFPT results. We find that the modes $\mathrm{A_{g}^1}$, $\mathrm{A_{g}^3}$, $\mathrm{B_{1g}^1}$, and $\mathrm{B_{1g}^3}$ arise primarily from nearly out-of-plane vibrations of Se atoms along the \textit{c}-axis, while the modes $\mathrm{A_{g}^2}$ and $\mathrm{B_{1g}^2}$ predominantly stem from in-plane vibrations along the \textit{b}- and \textit{a}-axes, respectively. Note that the $\mathrm{A_{g}}$ and $\mathrm{B_{1g}}$ modes exhibit distinctive responses to laser polarization as a result of their different symmetry. $\mathrm{B_{1g}}$ modes are detected mainly in the cross-configuration, whereas the $\mathrm{A_{g}}$ modes are predominantly visible in the parallel configuration, allowing for their clear differentiation. Furthermore, our experimental resolution enables us to discern between the $\mathrm{A^1_{g}}$ and $\mathrm{B^1_{1g}}$ modes, thus clarifying what was reported in previous Raman studies in which a mixed mode was detected \cite{oyedele2017pdse2,PhysRevB.26.6554, wang2012electronics}. To understand the anisotropic optical properties, we investigated the dependence of the Raman scattering versus the polarization angle and crystal orientation. The results are reported in the inset of Figure \ref{Fig1}(a). Under parallel conditions, the $\mathrm{A^1_{g}}$ mode intensity has maxima at $90^\circ$ and $270^\circ$, with a periodicity of $180^\circ$. The Raman intensity at a given angle $\theta$ can be fitted using curves of the form $\delta_{(\theta)} = \delta_a \cos^2(\theta + \phi) + \delta_b \sin^2(\theta + \phi)$, where $\delta_a$ or $\delta_b$ represents the Raman intensity along the $a$ and $b$ crystal axes, and $\phi$ is the fitting parameter \cite{yu2020direct}. This reveals that the $\mathrm{A_{g}}$ modes exhibit two-fold symmetry, aligning with the inherent in-plane anisotropy of the structure of PdSe$_2$, where the \textit{a}- and \textit{b}-axes correspond to high and low Raman intensity of the $\mathrm{A_{g}}$ modes, respectively. The symmetry of the Raman modes is analyzed and the results are provided in the Supplementary Information.
Figure \ref{Fig2}(a) presents the Raman spectra of bulk PdSe$_2$ measured in parallel configuration, where the polarization of the incident light is taken along the a-axis of the crystal, as a function of temperature. Temperature variations induce noticeable alterations in the frequency and intensity of the three Raman-active phonon modes. All observed Raman peaks experience a blue shift, attributed to lattice thermal anharmonicity, which contains contributions from phonon-phonon scattering and volume thermal expansion effects.
Interestingly, as the temperature decreases from 300 to 5 K, a prominent feature emerges, the clear visibility of the $\si{B^1_{1g}}$ mode on the high-frequency side of the $\si{A^1_{g}}$ which is highlighted in Figure \ref{Fig2}(b). 
At 120 K, further changes become evident, with the peak $\mathrm{A^1_{g}}$ reaching its maximum intensity, while modes $\mathrm{A^2_{g}}$ and $\mathrm{B^2_{1g}}$ are suppressed. In this temperature range, the intensity of the Raman mode with a higher frequency $\mathrm{A^3_{g}}$ remains relatively constant. Additionally, a marked transformation in the Raman spectrum is observed at higher frequencies when the temperature decreases from 250 to 5 K. In particular, the Raman mode at 273 cm$^{-1}$, associated with the $\mathrm{B^3_{1g}}$ mode, appears in the parallel configuration similar to the $\mathrm{B^1_{1g}}$ mode, which is not normally allowed. This suggests either a subtle change in the symmetry or a resonant Raman scattering process as observed in the breakdown of polarization Raman selection rules in few-layer TMDs by resonant Raman spectroscopy \cite{tan2021breakdown} .\\
To gain deeper insight into the anomalous phonon response of the $\si{A^1_{g}}$, $\si{A^2_{g}}$, $\si{A^3_{g}}$, $\si{B^1_{1g}}$ and $\si{B^3_{1g}}$ modes, we performed a Lorentzian lineshape fitting of the Raman spectra. In the following, we present the temperature dependence of Raman frequency, and full width at half maximum (FWHM) of each mode, while later we discuss the scattering strength obtained by the fittings. As shown in Figure \ref{Fig2}(c), for $T$ $>$ 100 K, a decrease of temperature results in a linear blueshift of all Raman frequencies. Conversely, in the low temperature range, all Raman mode frequencies tend to reach constant values.
The pronounced nonlinearity of the mode frequencies originates from the anharmonic phonon interactions \cite{balkanski1983anharmonic}. Taking three- and four-phonon scattering processes into account, corresponding to cubic and quartic anharmonicities, the temperature dependence of the Raman mode frequencies can be described by the following relation:
\begin{equation} \label{Eq2}
\begin{split}
\omega(T) &= \omega_0 + A \left(1+\frac{2}{e^x-1}\right) \\
&\quad + B \left(1+\frac{2}{e^y-1}+ \frac{3}{(e^y-1)^2} \right)
\end{split}
\end{equation}
Here, $\omega$ is the peak frequency which depend on temperature $T$, $\omega_{0}$ is the bare phonon frequency at $T$=0 K, $x = \frac{\hbar\omega_{0}}{2k_{b}T}$, $y = \frac{\hbar\omega_{0}}{3k_{b}T}$, while $A$ and $B$ are the anharmonic constants for the three- and four-phonon processes, respectively\cite{balkanski1983anharmonic}.\\
At high temperatures, when considering only the first terms of a Taylor expansion, Eq.\ref{Eq2} tends toward a linear dependence. The fit of Eq.\ref{Eq2} to the Raman frequencies is presented in Figure~\ref{Fig2}(c) (solid lines). The behavior of the $\mathrm{B_{1g}}$ phonon modes is taken from the close-to-cross configuration where all phonon modes are observed simultaneously and is presented in the Supplementary Information.  We also performed temperature-dependent Raman measurements on a trilayer $\si{PdSe_{2}}$ for comparison (see Supplementary Information). The fitted values of the anharmonic constants are presented in Table~I. These coefficients are lower than those observed in bulk PdSe$_2$.

\begin{table}[h]
\caption{\footnotesize The values of anharmonic constants obtained from the temperature-dependent analysis of the Raman mode frequencies of bulk and trilayer PdSe$_2$ samples.}

\begin{center} {\footnotesize
\begin{tabular}{p{2cm} p{2cm} p{2cm} p{2cm}}
\hline
\hline 
\\
mode & $\omega_{0}$ ($\si{cm^{-1}})$ & A ($\si{cm^{-1}})$ & B ($\si{cm^{-1}})$ \\
\hline
\hline
bulk 
\\
\hline
$\si{A_{g}^1}$    & 148.6392 & -0.3853 & -0.0443 \\
$\si{A_{g}^2}$    & 215.2802 & -2.3291 & -0.2067 \\
$\si{A_{g}^3}$    & 267.1185 & -2.3042 & -0.2727 \\
$\si{B_{1g}^1}$   & 153.4026 & -1.1921 & -0.0245 \\
$\si{B_{1g}^2}$   & 231.5289 & -2.4889 & -0.0762 \\
$\si{B_{1g}^3}$   & 277.1462 & -2.0256 & -0.5453 \\
\hline
trilayer\\
\hline
$\si{A_{g}^1}$    & 152.7897 & -0.5447 & -0.00134  \\
$\si{B_{1g}^2}$   & 233.6546 & -1.2792 & -0.00509 \\
\hline
\end{tabular} }
\end{center}
\label{turns}
\end{table}
The ratio \(A/B\) is high for phonons \(\mathrm{A_{g}}\) and $\mathrm{B_{1g}}$ modes, due to the higher probability of decay process of optical phonons into two acoustic phonons than the three acoustic phonons. Additionally, the value of the A coefficient for $\si{A_{g}^1}$ is the smallest among all the modes. A similar non-linear temperature dependence of the Raman mode frequencies was also observed for other materials such as black phosphorus, $\si{MoS_{2}}$, and $\si{SnSe_{2}}$ nanosheets \cite{lapińska2016temperature, taube2014temperature, taube2015temperature, kong2015raman, taube2016temperature}.\\
The behavior of Raman modes widths mirrors that of their frequencies as the temperature decreases: they decrease initially and then tend to saturate at temperatures below 100 K (see the Supplementary Information).
Concerning the scattering strength of the phonon modes, Figure \ref{Fig3A} shows the temperature-dependent spectral intensity of the $\si{A_{g}^1}$, $\si{A_{g}^2}$, and $\si{A^3_{g}}$ modes. Specifically, $\si{A_{g}^1}$ reaches its maximum intensity at 120 K, while $\si{A^2_{g}}$ exhibits a zero intensity at the same temperature, followed by partial recovery. We note that this behavior was also observed for the $\si{B_{1g}}$ modes in the close-to-cross-polarization configuration (see the Supplementary Information). 

To further explore the anomalous behavior of the $\si{A_{g}^1}$ mode, we performed temperature-dependent anti-Stokes Raman measurements, as reported in Figure \ref{Fig3}(a). We find that the $\si{A_{g}^1}$ mode exhibits a gradual increase in intensity, reaching its maximum at about 90 K, mirroring the behavior observed in the Stokes measurements. In contrast, the $\si{A_{g}^3}$ modes showed a consistent reduction in intensity with decreasing temperature, following the typical temperature-dependent behavior and ultimately becoming undetectable at 5 K, as shown in Figure \ref{Fig3}(b).
\begin{figure*} 
\includegraphics[width=0.8\textwidth]{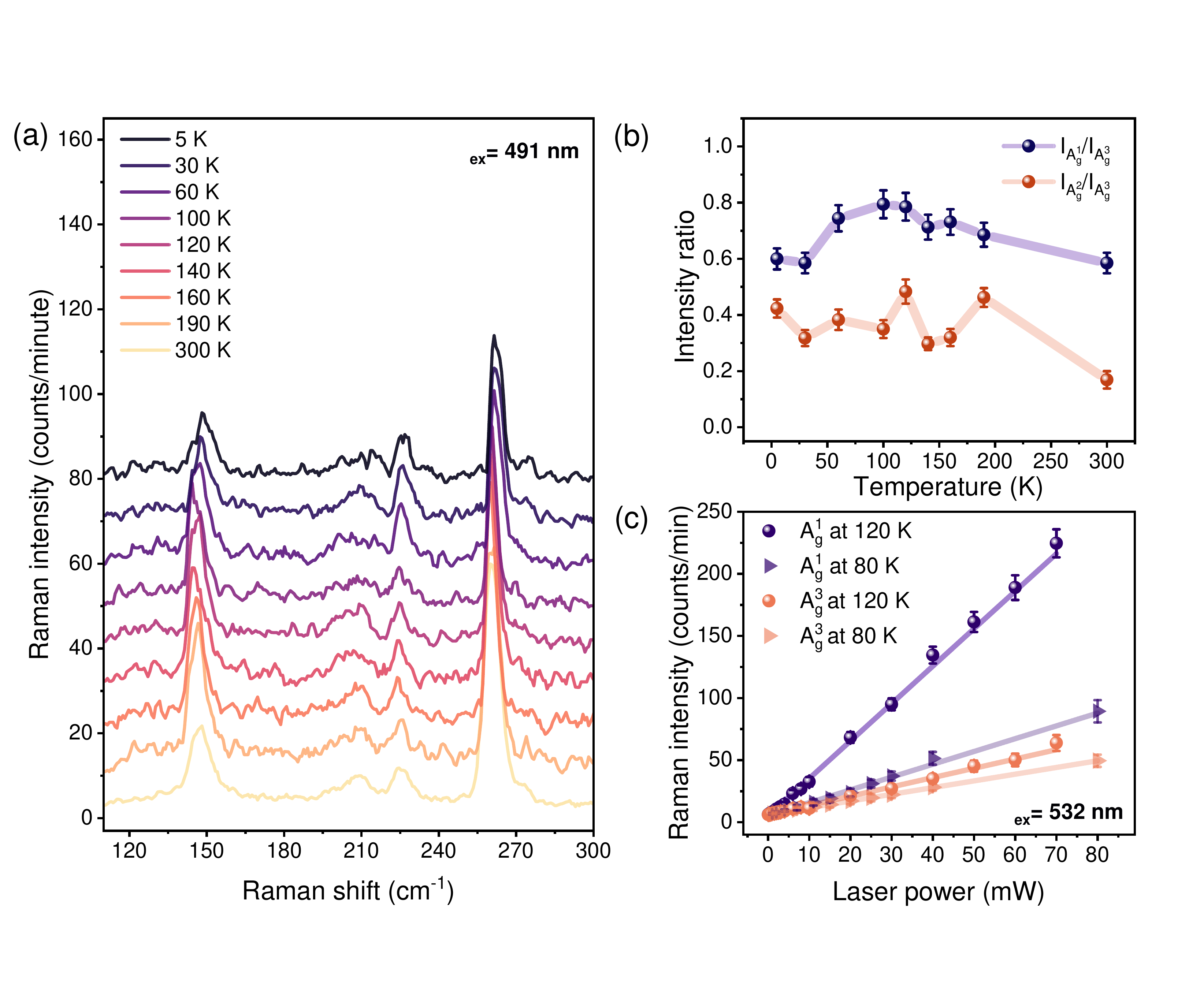}
\caption{(a) Temperature-dependent Raman spectra of bulk $\si{PdSe_{2}}$ under parallel configuration (xx) with an excitation energy of 2.52 eV (491 nm). (b) The intensity ratio of the Raman $\si{A^1_{g}}$ and $\si{A^2_{g}}$ to $\si{A^3_{g}}$ mode. (c) Power dependence of the Raman intensities of the $\si{A^1_{g}}$ and $\si{A^3_{g}}$ phonon modes for fixed temperatures of 120 and 80 K. The slope values of the $\si{A^1_{g}}$ and $\si{A^3_{g}}$ at 120 K and 80 K are 3.19 c/m.mW, 1.11 c/m.mW, 0.76 c/m.mW, and 0.54 c/m.mW, respectively. }
\label{Fig4}
\end{figure*}
\begin{figure} [t!]
\includegraphics[width=\columnwidth]{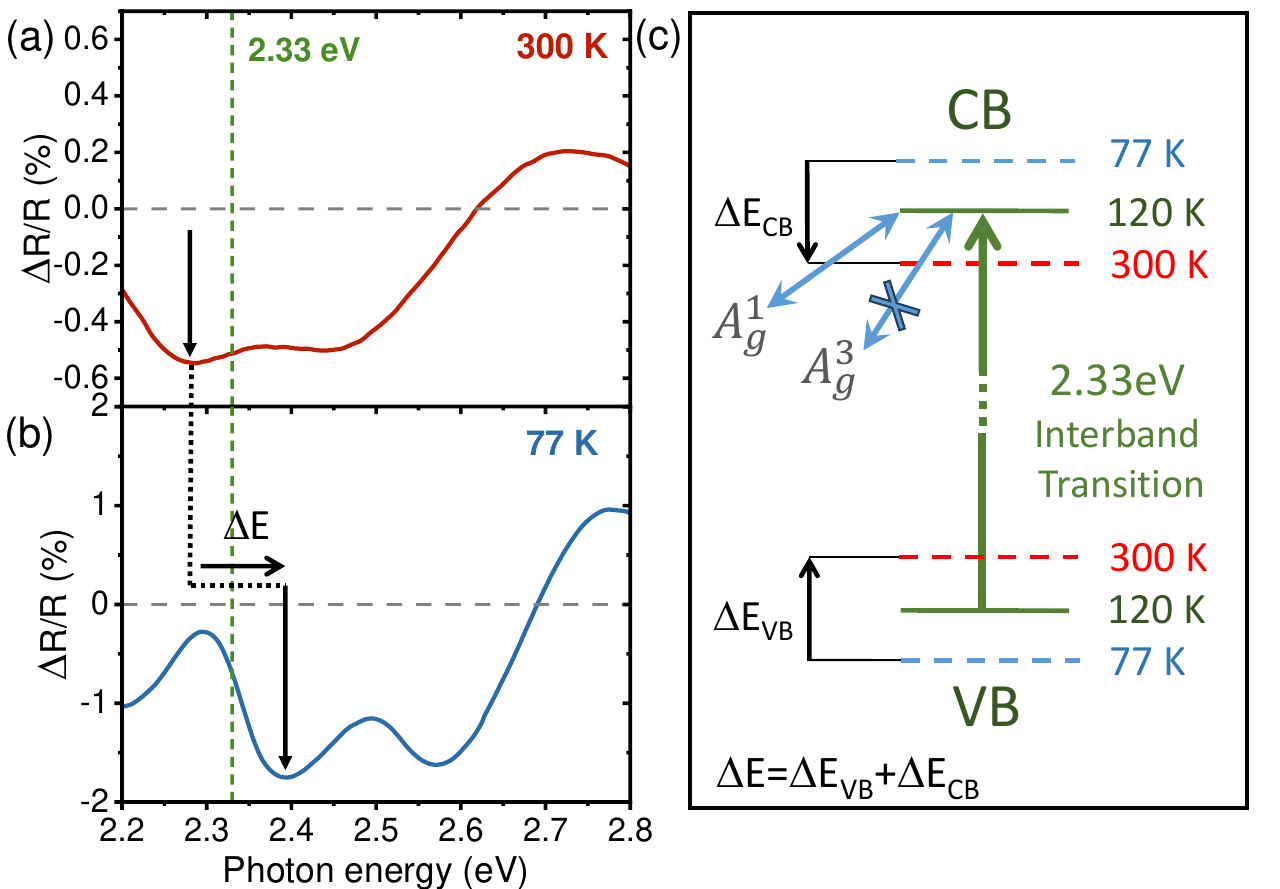}
\caption{Transient reflectivity measurements of PdSe$_2$ measured at (a) 300 K and (b) 77 K showing the differential reflectivity, $\Delta R/R$ at 1 picosecond time delay after photoexcitation. The arrows indicate the shift of the transition around 2.33 eV toward higher energies as the sample temperature is reduced. (c) A schematic representation of the variation in some interband optical transitions, showing how resonant excitation at 120 K interacts differently with various phonon modes.}
\label{Fig5}
\end{figure}
\begin{figure*}
\centering
\includegraphics[width=0.7\textwidth]{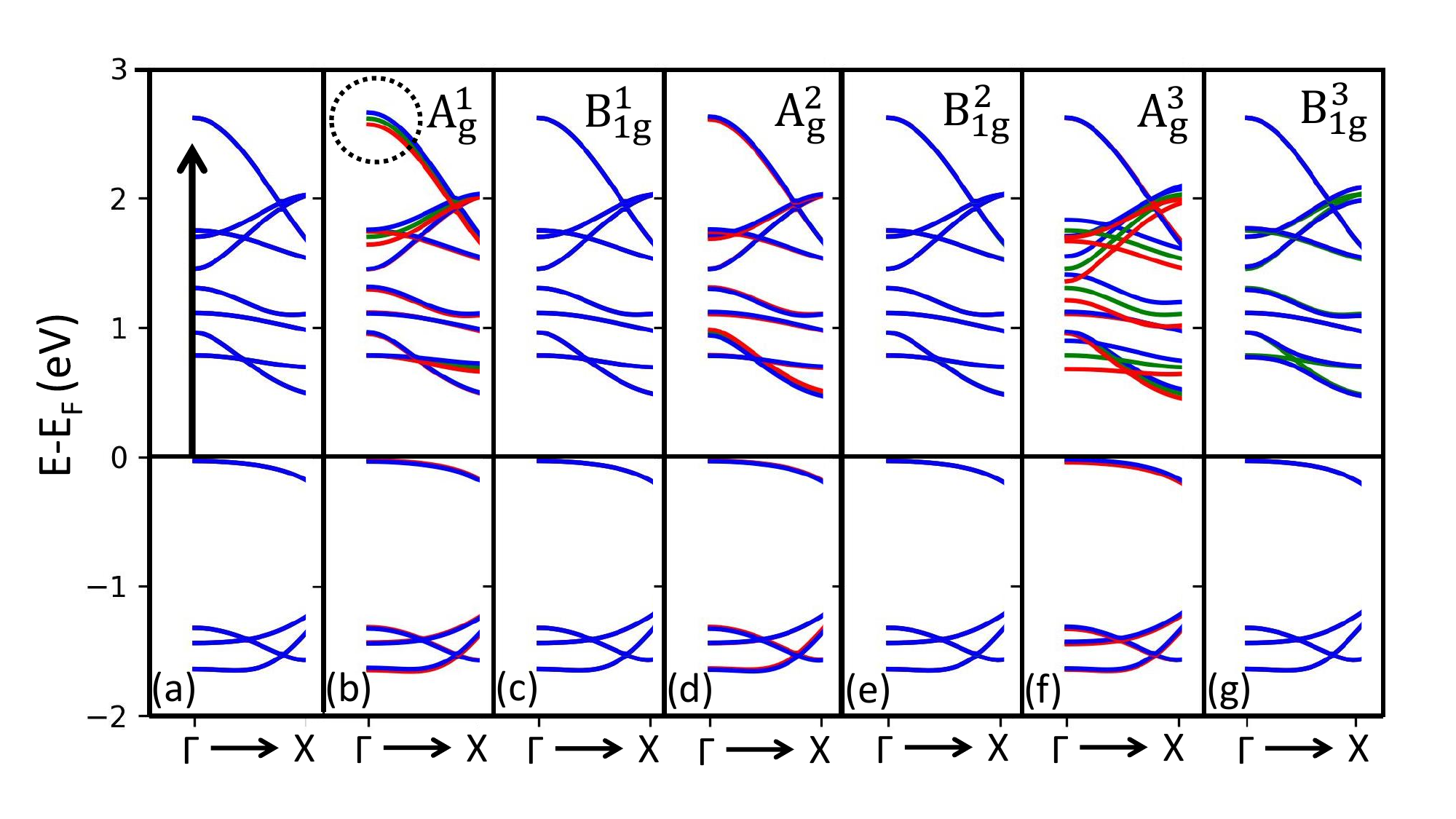}
\caption{Estimating the electron-phonon coupling using DFT calculations for the electronic dispersion of bulk PdSe$_2$ considering unperturbed (green lines) (a) and different frozen phonons (blue and red for opposite displacements) (b) $\si{A^{1}_{g}}$, (c) $\si{B^{1}_{1g}}$, (d)  $\si{A^{2}_{g}}$, (e) $\si{B^{2}_{1g}}$, (f) $\si{A^{3}_{g}}$, and (g)  $\si{B^{3}_{1g}}$.}
\label{Fig6}
\end{figure*}
The anti-Stokes to Stokes scattering ratio is a useful quantity for determining the effective phonon temperature, as discussed in Ref. \cite{compaan1984resonance}. In the absence of resonant conditions, the function $\phi$ can give an indication of the phonon temperature $\phi\left(\frac{I_{S}}{I_{AS}}\right) = -\frac{\hbar\Omega}{k_b \ln\left(\frac{I_{S}}{I_{AS}}\right)}
$, where $\Omega$ is the phonon frequency. Figure \ref{Fig3}(c) presents the values of $\phi\left(\frac{I_{S}}{I_{AS}}\right)$ determined for the  $\si{A_{g}^1}$ and $\si{A_{g}^3}$ phonon modes at different temperatures. $\phi\left(\frac{I_{S}}{I_{AS}}\right)$  can be used to extract the temperature of the $\si{A_{g}^3}$ phonon, as it demonstrates a linear behavior when the sample temperature is reduced. However, for $\si{A_{g}^1}$, this relationship does not exhibit the expected linear behavior, suggesting a possible resonant excitation effect linked to $\si{A_{g}^1}$.\\
To clarify the effect of resonance enhancement of the $\si{A^1_{g}}$ mode at 120 K, we measured the temperature-dependent Raman spectra of bulk PdSe$_2$ employing alternative excitation energy. If the anomalous increase of $\si{A^1_{g}}$ mode is due to resonance with 532 nm excitation, we anticipate that it will not be visible when using other laser lines. Figure \ref{Fig4}(a) shows the results of Raman measurements detected using 491 nm (2.52 eV) excitation in parallel configuration. As the temperature decreases from 300 K to 5 K, all phonon modes exhibit a blue shift, consistent with the observations from the 2.33 eV excitation laser measurements. However, the Raman spectrum is relatively unchanged, and the subtle change in symmetry observed using 2.33 eV excitation is entirely absent. The absence of phonon-selective resonant enhancement is discernible through the negligible change observed in the intensity ratios between $\si{A^{1}_{g}}$ to $\si{A^{3}_{g}}$ and $\si{A^{2}_{g}}$ to $\si{A^{3}_{g}}$ as shown in Figure \ref{Fig4}(b).\\
By altering the excitation energy, we observed that the peculiar temperature-dependent amplitude behaviour was absent. We then investigated whether changing the energy of the electronic states had a similar effect on the anomaly. The thickness of PdSe$_2$ has a major influence on its electronic band structure and optical resonances. To further explore this effect and understand the dependence of this behaviour on the flake thickness, we studied the temperature-dependent Raman spectra of a mechanically exfoliated thin sample (trilayer). The electronic band gap of PdSe$_2$ has been reported to increase significantly with the transition from the bulk to the thin-layer regime, causing a substantial modification of the entire electronic landscape \cite{oyedele2017pdse2}. We observe significant frequency shifts of several Raman modes, ranging from 5 to 9 cm$^{-1}$, as we move from the bulk to the trilayer system, which is in excellent agreement with the other reports \cite{oyedele2017pdse2,yu2021giant} (see the Supplementary Information). This large blueshift from bulk to trilayer is attributed to the strong interlayer coupling and the change of the in-plane lattice constants.
Temperature-dependent measurements revealed that the resonance effect is not as strong in the trilayer system, and the significant increase of $\si{A^{1}_{g}}$ that was seen in the bulk is completely absent, which is in agreement with our interpretation.\\
Using 2.33 eV excitation, we investigated the power dependent behavior of the $\si{A^{1}_{g}}$ and $\si{A^{3}_{g}}$ modes at selected temperatures of 120 K and 80 K aiming to assess the influence of optical excitation on these modes under resonant (120 K) and non-resonant (80 K) conditions. As shown in Figure \ref{Fig4}(c), we clearly see that the $\si{A^{3}_{g}}$ mode exhibits a similar power-dependent behavior at both temperatures, i.e. there is no significant resonance effect, which is almost identical to the behavior of $\si{A^{1}_{g}}$ at 80 K. In contrast, at 120 K,  $\si{A^{1}_{g}}$ exhibits a remarkable sensitivity to increasing laser power, indicating a resonant effect at this temperature. This effect can be attributed to an enhancement of the prefactor in the Raman response associated with resonant optical properties at 120 K. Consequently, all experimental observations consistently point towards a resonant Raman effect. The non-resonant components within the Raman spectrum may neutralize the resonant Raman component, resulting in the suppression of the Raman signal, as explained in \cite{ralston1970resonant}. This phenomenon accounts for the observed extinction of the in-plane $\si{A^{2}_{g}}$ and $\si{B^{2}_{g}}$ modes at 120 K.\\
To gain experimental insights into the resonance phenomena in the Raman response of bulk PdSe$_2$ and to explore its dependence on temperature, we employed optical spectroscopy within the range of Raman excitation energy. We used broadband transient reflectivity experiments that involved photoexciting the sample with an optical pump pulse centered at approximately 2.30 eV and subsequently measuring the differential reflectivity, $\si{\Delta R/R}$, after a 1 picosecond time delay. Figure \ref{Fig5}(a) shows the 1 ps optical response at room temperature where a negative feature of $\si{\Delta R/R}$  approximately at 2.28 eV (indicated by arrows) is identified. Upon reducing the temperature to 77 K, we observed a discernible shift of the reflectivity peak towards higher energies (about 2.39 eV), as shown in Figure \ref{Fig5}(b). Notably, this transition sweeps the 2.33 eV energy (dashed green line, the Raman excitation energy) as the temperature was lowered from 300 K to 77 K. Analogous behavior was also observed in other materials such as $\si{WSe_2}$ and $\si{MoSe_2}$ \cite{yan2014photoluminescence, ye2018nonlinear}. Figure \ref{Fig5}(c) illustrates this transition and energy shift, resembling a resonance at 2.33 eV, possibly at 120 K. This observation supports the conjecture of a significant change in Raman modes on the basis of electronic transitions. A recent study has revealed that the transient absorption spectrum of an 8-layer PdSe$_2$ displays coherent oscillations at a frequency of 143 cm$^{-1}$,  identified as the only optical phonon in the transient optical response, suggesting the strong interaction between the $\si{A^{1}_{g}}$ mode and the electronic states \cite{li2021phonon}. Given that the excitonic characteristics of the 2.31 eV optical transition are discussed in \cite{li2021phonon}, it is reasonable to attribute the observed Raman resonance to strong exciton-phonon interactions, similar to those seen in resonant excitation in monolayers of WS$_2$ and WSe$_2$ \cite{del2016atypical,mcdonnell2018probing}. 
To investigate the electron-phonon interactions, we conducted theoretical calculations based on DFT.  We calculated the band structure of bulk PdSe$_2$ perturbed by small displacements (both positive and negative) of the atoms according to each of the $\Gamma$ point phonon eigenvectors. This highlights in a simple way which optical transitions should give rise to strong phonon scattering via the deformation potential.  
The results are presented in Figures \ref{Fig6}(a to g), revealing not only that the electron-phonon coupling is different for the $\si{A^{1}_{g}}$ and $\si{A^{3}_{g}}$ modes, but that they also couple to different electronic states. Specifically, the bands located approximately around 2.5 eV above the Fermi level display strong coupling with $\si{A^{1}_{g}}$, although the overall electron-phonon coupling is more pronounced for $\si{A^{3}_{g}}$ (we show the calculated strengths of the electron-phonon coupling for all the Raman modes active in backscattering in the Supplementary Information). Hence, through resonant optical excitation, the phonon behavior of PdSe$_2$ can be manipulated due to the strong and diverse electron-phonon interactions specific to each phonon mode and electronic state. These results illustrate the significant potential of resonance Raman scattering in examining layered TMDs, particularly in highly anisotropic PdSe$_2$. 

\section{CONCLUSION\label{CONCLUSIONS}}
We examined phonon behavior in bulk PdSe$_2$ using Raman spectroscopy over a temperature range of 5 to 300 K. We found that there is a temperature-dependent enhancement of specific Raman modes linked to resonance effects with particular optical transitions. For 2.33 eV excitation, we observed an opposite response: a significant increase in the intensity of the out-of-plane $\si{A^{1}_{g}}$ mode, which peaks at 120 K, while there was a corresponding decrease in the in-plane $\si{A^{2}_{g}}$ and $\si{B^{2}_{g}}$ modes, while the out-of-plane $\si{A^{3}_{g}}$ mode remains almost unchanged. The interpretation of the experimental results is supported by density functional theory (DFT) calculations, which confirm an enhanced electron-phonon coupling strength related to relevant optical transition energies. The anomalous behavior at 120 K can be explained by electron-phonon coupling and resonance effect in PdSe$_2$. Additionally, we observed nonlinear frequency shifts in all phonon modes, originating from multiple phonon scattering decay channels. This work provides essential insights into the low-temperature electronic and vibrational properties of this anisotropic material, offering valuable information for further applications of the thermal characteristics of PdSe$_2$ in optoelectronic devices.  Our study serves to inspire further experimental and theoretical advancements in understanding the physics of two-dimensional materials.

\section{METHODS \label{methods}}
\subsection{Sample preparation and characterization:} 
The bulk PdSe$_2$ crystals were purchased from HQ Graphene \cite{HQGraphene}. PdSe$_2$ thin flakes were mechanically exfoliated from the bulk crystals using Nitto SPV 224R tape and then transferred to a substrate of 280 nm SiO$_2$/Si substrate.
An optical microscope was used to determine and identify the thinner flakes based on their optical contrast. Using a Veeco Dimension 3000 atomic force microscope (AFM) the thickness of the trilayer was determined.\\
\subsection{Experimental procedures and measurement technique:} The Raman spectra of bulk and trilayer PdSe$_2$ samples were measured, using a micro-Raman system in backscattering geometry. To prevent any significant local temperature increase in the sample, an appropriate laser power was employed during the experiment. Two lasers with excitation energies of 2.33 eV (532 nm) and 2.52 eV (491 nm) were used. All measurements were performed in a vacuum using an optical coldﬁnger cryostat (Janis ST500) with a working distance (about 2 mm). The laser beam was focused to a spot size of 1 $\mu$m onto the sample by a 50× microscope objective lens and a long working distance of 12 mm. The laser power was set to a maximum of 100 $\si{\mu W}$. The backscattered light, collected by the same objective lens, was transmitted and directed onto the spectrometer's entrance slit. Following this path, the scattered signal underwent dispersion within the spectrometer and was subsequently detected using a liquid nitrogen-cooled back-illuminated charge-coupled-device (CCD) detector, specifically a low-etaloning PyLoN:1000BR\textunderscore eXcelon, featuring a 1340 $\times$ 100 pixels CCD.\\
Polarization-resolved Raman measurements incorporated an analyzer into the light path detection to determine the scattered light's polarization direction. Throughout the measurements, the PdSe$_2$ samples remained fixed on the sample stage, while the polarization directions of the incident and scattered light were manipulated using a halfwave plate and analyser in steps of 15 degrees. The Raman spectra were obtained under both parallel and cross configurations. In the parallel configuration (xx), the scattered light's polarization aligned parallel to the incident light, while in the cross configuration (xy), the polarization was perpendicular. This distinction was achieved using the same analyzer.\\
\subsection{Phonon displacements and DFT calculations:} 
Phonon displacements and eigenmodes were computed using VASP \cite{kresse1999ultrasoft, hafner2008ab} by evaluating force constants for a 2$\times$2$\times$2 supercell through the frozen phonon approach implemented in phonopy. The structures were relaxed in VASP to achieve forces below $10^{-6}$ eV/\AA, with a kinetic energy cutoff of 300 eV and a Monkhorst-Pack k-point density of 8$\times$8$\times$6 for the primitive unit cell \cite{monkhorst1976special}. In the DFT calculations, we employed the regularized SCAN meta-GGA exchange-correlation functional \cite{bartok2019regularized}, along with the rVV10 kernel for the inclusion of van der Waals forces between layers \cite{sabatini2013nonlocal} (more details can be found in the Supplementary Information).


\section*{DATA AVAILABILITY}
The experimental data are accessible through:
\href{https://doi.org/10.5281/zenodo.10424154}{https://doi.org/10.5281/zenodo.10424154}\\
Inputs to the computational codes are available free of charge at \href{https://doi.org/10.15125/BATH-01356}{https://doi.org/10.15125/BATH-01356} 
\def\bibsection{\section*{~\refname}} 
\bibliography{bibliography}

\section*{ACKNOWLEDGEMENTS\label{ack}}
The authors acknowledge financial support by the Deutsche Forschungsgemeinschaft (DFG) through project German Research Foundation via project No. 277146847 - CRC 1238: Control and Dynamics of quantum Materials. Computational work was performed on the University of Bath’s High Performance Computing Facility and was supported by the EU Horizon 2020 OCRE/GEANT project “Cloud funding for research” .
\section*{AUTHOR CONTRIBUTIONS}
O.A. conducted the Raman measurements and analyzed the data. The study was supervised by H.H. and P.v.L. D.W. developed the theoretical support in discussions with E.C. and conducted the calculations. C.S. performed the optical spectroscopy experiments. O.A. wrote the manuscript with the support of H.H. The results were discussed and the paper was reviewed by O.A., D.W., C.S., E.C., F.P., H.H., and P.v.L.

\section*{COMPETING INTERESTS}
The authors declare no competing financial or non-financial interests.
\section*{ADDITIONAL INFORMATION}
\textbf{Supplementary Information} The online version contains supplementary material
available at https://doi.org/XXX.

\end{document}